\documentclass[aps,prl,showpacs,preprint]{revtex4} 
\usepackage{graphicx}
\usepackage{amssymb}
\usepackage{epsfig}

\begin{document}

\title{Low Dimensional Behavior of Large Systems of Globally\\ Coupled Oscillators}

\author{Edward Ott and Thomas M. Antonsen}

\affiliation{University of Maryland, College Park, MD 20742}


\begin{abstract}
It is shown that, in the infinite size limit, certain systems of
globally coupled phase oscillators display low dimensional
dynamics. In particular, we derive an explicit finite set of
nonlinear ordinary differential equations for the macroscopic
evolution of the systems considered. For example, an exact, closed
form solution for the nonlinear time evolution of the Kuramoto
problem with a Lorentzian oscillator frequency distribution
function is obtained. Low dimensional behavior is also
demonstrated for several prototypical extensions of the Kuramoto
model, and time-delayed coupling is also considered.
\end{abstract}
\pacs{05.45.Xt, 05.45.-a, 89.75.-k}
 \maketitle

 {\bf Because synchronous behavior in large groups consisting of many coupled
 oscillators has been widely observed in many situations, the behavior of
 such systems has long been of interest. Since the problem is
 difficult to solve in general, much work has been done on the
 simple paradigmatic case of globally coupled phase oscillators.
 Even in this simple context, however, much remains unclear,
 particularly when considering situations in which a large oscillator
 population interacts with external dynamical systems, or when
 there are communities of interacting oscillators with different
 community
 and connection characteristics, etc. In this paper we consider an
 approach that allows the study of the time evolving dynamical
 behavior of these types of systems by an exact reduction to a
 small number of ordinary differential equations. This reduction
 is achieved by considering a restricted class of states. In spite
 of this restriction, for at least one significant example [see
 preceding article], consideration of our derived ordinary
 differential equations appears to yield dynamics in precise
 agreement with results obtained from considerations not imposing
 this restriction. Thus we believe that our results may be useful
 in many other contexts.}
 \section{I. Introduction}
Understanding the generic behavior of systems consisting of large
numbers of coupled oscillators is of great interest because such
systems occur in a wide variety of significant
applications\cite{Pikovsky}. Examples are the synchronous flashing
of groups of fireflies, coordination of oscillatory neurons
governing circadian rhythms in animals\cite{Yamaguchi},
entrainment in coupled oscillatory chemically reacting
cells\cite{Kiss}, Josephson junction circuits\cite{Wiesenfeld},
neutrino oscillations\cite{Pantaleone}, bubbly
fluids\cite{Haubler}, etc. A key contribution in this area was the
introduction of the following model by Kuramoto\cite{Kuramoto},
\begin{equation}
d\theta _i(t)/dt=\omega _i+\frac{K}{N} \sum ^N_{j=1}\sin (\theta
_j(t)-\theta _i(t)) \ ,
\end{equation}
where the state of oscillator $i$ is given by its phase $\theta
_i(t)$, $(i=1,2,\ldots,N)$, $\omega _i$ is the natural frequency
of oscillator $i$, and the coupling constant $K$ specifies the
strength of the influence of one oscillator on another. It has
been shown\cite{Kuramoto,model} that in the $N\rightarrow \infty$
limit there is a continuous phase transition such that, for $K$
below a critical value $(K<K_c)$, no coherent behavior of the
system occurs (i.e., there is no global correlation between the
oscillator phases), while above the critical coupling strength
$(K>K_c)$, the system displays global cooperative behavior (i.e.,
partial or complete synchronization of the phases).

Among other problems related to (1) that we shall also consider
are the case where there is a sinusoidal periodic external drive
of strength $\Lambda $ added to the righthand side of (1) (see
Refs.\cite{Sakaguchi} and \cite{Antonsen}),
\begin{equation}
d\theta _i/dt=\omega _i+\frac{K}{N}\sum ^N_{j=1}\sin (\theta
_j-\theta _i)+\Lambda \sin (\Omega t-\theta _i) \ ,
\end{equation}
and the case where there are several communities of different
kinds of oscillators where the evolution of the phases $\theta
^\sigma _i(t)$ of oscillators in community $\sigma $ is given by
(see Refs.\cite{Barreto,Montbrio})
\begin{equation}
d\theta _i^\sigma /dt=\omega _i^\sigma + \sum ^s_{\sigma
'=1}\frac{K_{\sigma\sigma '}}{N_{\sigma '}}\sum _{j=1}^{N_{\sigma
'}}\sin (\theta _j^{\sigma '}-\theta _i^\sigma ) \ .
\end{equation}
Here $\sigma =1,2,\ldots $,$s$, $N_\sigma $ is the number of
oscillators of type $\sigma $, and $K_{\sigma \sigma '}$ is the
strength of the coupling from oscillators in community $\sigma '$
to oscillators in community $\sigma $. For all three cases
(Eqs.~(1), (2), (3)), we are interested in the limit $N\rightarrow
\infty $. We will also consider such problems with time delayed
coupling (e.g., $\theta _j(t)\rightarrow \theta _j(t-\tau )$ in
Eqs.~(1)--(3)).

 The problem stated in Eq.~(2) was first
considered by Sakaguchi\cite{Sakaguchi}.  It can, for example, be
motivated as a model of circadian rhythm\cite{Yamaguchi}.
Circadian rhythm in mammals is governed by the suprachiasmatic
nucleus that is located in the brain and consists of a large
population of oscillatory neurons. These neurons presumably couple
with each other and are also influenced (though the optic nerve)
by the daily variation of sunlight (modeled by the term in (2)
involving $\Lambda $).  In \cite{Antonsen}, we found numerical and
analytical evidence that the bifurcations and macroscopic dynamics
of (2) with large $N$ appeared to be similar to what might be
expected for the dynamics of a two dimensional dynamical system.
This observation was the motivation for the present paper.

The problem stated in Eq.~(3) has been previously considered in
Refs.\cite{Barreto} and \cite{Montbrio} where the linear stability
of the incoherent state was investigated along with numerical
solutions for the nonlinear evolution.

\section{II. Nature of the Main Result}
Considering the limit $N\rightarrow \infty$, the state of the
oscillator system at time $t$ can be described by a continuous
distribution function, $f(\omega ,\theta ,t)$, in frequency
$\omega $ and phase $\theta $ for the problems in Eqs.~(1) and (2)
or by $f^\sigma (\omega ,\theta ,t)$ with $\sigma =1,2,\ldots ,s$
for the problem in Eq.~(3), where
\[
\int ^{2\pi }_0f(\omega ,\theta ,t)d\theta =g(\omega ) \ \ {\rm
or} \ \ \int ^{2\pi}_0f^\sigma (\omega ,\theta ,t)d\theta
=g^\sigma (\omega ) \ ,
\]
and $g(\omega )$ and $g^\sigma (\omega )$ are time independent
oscillator frequency distributions.

Our main result is as follows.  For initial distribution functions
$f(\omega ,\theta ,0)$ satisfying a certain set of conditions that
we will specify later in this paper, we show that
\begin{enumerate}
\item [(i)] the evolution of $f(\omega ,\theta ,t)$ from $f(\omega
,\theta ,0)$ continues to satisfy the specified conditions,

\item [(ii)] for appropriate $g(\omega )$ [or $g^\sigma (\omega
)$], the macroscopic system state obeys a finite set of nonlinear
ordinary differential equations, which we obtain explicitly.
\end{enumerate}

Concerning (i), we define a distribution function $h(\omega
,\theta )$ as a function for which $h\geq 0$ and $\int _0^{2\pi
}d\theta \int d\omega h=1$, and the distribution functions
$h(\omega ,\theta )$ satisfying our conditions form a manifold $M$
in the space $D$ of all possible distribution functions. What
point (i) says is that initial states in $M\subset D$ evolve to
other states in $M$. Thus $M$ is `invariant' under the dynamics.
Concerning point (ii), we use the so-called `order-parameter'
description to define the macroscopic system state. We define the
order parameter (or parameters in the case of Eq.~(3))
subsequently (Eq.~(5)) in terms of an integral over the
distribution function $f$ (or $f^\sigma$ for (3)), where this
order-parameter integral globally quantifies the degree to which
the entire ensemble of oscillators (or ensembles $\sigma $ for
(3)) behaves in a coherent manner.  According to point (ii) the
evolution of the order parameters is exactly finite dimensional
even though the manifold $M$ and the dynamics of the distribution
function $f$ as it evolves in $M$ are infinite dimensional.

The macroscopic dynamics we obtain allows for much simplified
investigation of the systems we study. For example, we obtain an
exact closed form solution for the nonlinear time evolution of the
Kuramoto problem, Eq.~(1), for the case of Lorentzian $g(\omega
)$. Our formulation will be practically useful if at least some of
the macroscopic order-parameter attractors and bifurcations of the
full dynamics in the space $D$ are replicated in $M$.  In this
regard, we note that numerical solutions of the system (2) for
large $N$ have been carried out in Ref.~\cite{Antonsen}, and the
resulting macroscopic order-parameter attractors, as well as their
bifurcations with variation of system parameters, have been fully
mapped out. Comparing these numerical results for the full system
(Eq.~(2)) with results for the corresponding low dimensional
system for the dynamics on $M$ (Eq.~(14)), we find that {\it all}
(not just some) of the macroscopic order-parameter attractors and
bifurcations of Eq.~(2) with Lorentzian $g(\omega )$ are precisely
and quantitatively captured by examination of the dynamics on $M$.
These results for the problem given by Eq.~(2) suggest that our
approach may be useful for other situations such as Eqs.~(3).
Another notable point is that Ref.~\cite{Antonsen} also reports
numerical simulation results for Eq.~(2) with large $N$ for the
case of a Gaussian oscillator distribution function, $g(\omega
)=(2\pi \Delta ^2)^{-1/2}\exp [-(\omega - \omega _0)^2/(2\Delta
^2)]$, and the macroscopic order-parameter attractors and
bifurcations in this case are found to be the same as those in the
Lorentzian case (albeit at different parameter values). Thus, at
least for problem (2), phenomena for Lorentzian $g(\omega )$ are
not special and should give a useful indication of what can be
expected for other unimodal distributions $g(\omega )$.

\section{III. Derivation for the Example of the Kuramoto Problem}
We now support points (i) and (ii) for the case of the Kuramoto
problem, Eq.~(1). Following that, we will consider other problems,
including those associated with Eqs.~(2) and (3). Because of its
relative simplicity, in this section we use the Kuramoto problem
as an example, but we emphasize that our interest is primarily in
developing a method that will be useful in less simple cases, such
as the problems stated in Eqs.~(2) and (3) (see Sec. IV).
Following Kuramoto\cite{Kuramoto,model}, we note that the
summation in Eq.~(1) can be written as
\[
\frac{1}{N}\sum _j\sin [\theta _j-\theta _i]=Im \left\{
e^{-i\theta _i}\frac{1}{N} \sum _je^{i\theta _j}\right\}
=Im[re^{-i\theta _i}] \ ,
\]
where $r=N^{-1}\sum \exp (i\theta _j)$. Letting $N\rightarrow
\infty$ in Eq.~(1), $f(\omega ,\theta ,t)$ satisfies the following
initial value problem,
\begin{equation}
\partial f/\partial t+\partial /\partial \theta \left\{ [\omega
+(K/2i)(re^{-i\theta }-r^*e^{i\theta })]f\right\}=0 \ ,
\end{equation}
\begin{equation}
r=\int ^{2\pi }_0d\theta \int ^{+\infty}_{-\infty }d\omega
fe^{i\theta } \ ,
\end{equation}
where $r(t)$ is the {\it order parameter}, and Eq.~(4) is the
continuity equation for the conservation of the number of
oscillators. Note that by its definition (5), $r$ satisfies
$|r|\leq 1$. Expanding $f(\omega ,\theta ,t)$ in a Fourier series
in $\theta $, we have
\[
f=(g(\omega )/2\pi )\left\{ 1+[\sum ^\infty _{n=1}f_n(\omega
,t)\exp (in\theta )+c.c.]\right\}\ , \] where $c.c.$ stands for
complex conjugate. We now consider a restricted class of
$f_n(\omega ,t)$ such that
\[
f_n(\omega ,t)=(\alpha (\omega ,t))^n\ ,
\]
 where $|\alpha (\omega
,t)|\leq 1$ to avoid divergence of the series. Substituting this
series expansion into Eqs.~(4) and (5), we find the remarkable
result that this special form of $f$ represents a solution to (4)
and (5) if
\begin{equation}
\partial \alpha /\partial t+(K/2)(r\alpha ^2-r^*)+i\omega \alpha =0 \ ,
\end{equation}
\begin{equation}
r^*=\int ^{+\infty}_{-\infty}d\omega \alpha (\omega ,t )g(\omega )
\ .
\end{equation}
Thus this special initial condition reduces the $\theta$-dependent
system, (4), (5) to a problem (6), (7) that is $\theta
$-independent. However, we emphasize that Eqs.~(6) and (7) still
constitute an infinite dimensional dynamical system becuase any
initial condition is a {\it function} of $\omega $, namely $\alpha
(\omega , 0)$. Performing the summation of the Fourier series
using $\sum ^{\infty}_{n=1}x^n=x/(1-x)$, we obtain
\begin{equation}
f(\omega ,\theta ,t)=\frac{g(\omega )}{2\pi } \frac{(1-|\alpha |
)(1+|\alpha | )}{(1-|\alpha | )^2+4|\alpha | \sin
^2[\frac{1}{2}(\theta -\psi )]} \ ,
\end{equation}
where $\alpha \equiv |\alpha | e^{-i\psi }$ and $\psi $ real. For
$|\alpha | <1$ we can explicitly verify from Eq.~(8) that $f\geq
0$, $\int d\theta f=g(\omega )/2\pi $, and that as $|\alpha |
\nearrow 1$ we have $f\rightarrow \delta (\theta -\psi )g(\omega
)/2\pi $. In order that Eqs.~(6) and (7) represent a solution of
Eq.~(5) for all finite time, we require that, as $\alpha (\omega
,t)$ evolves under Eqs.~(6) and (7) that $|\alpha (\omega ,t)|\leq
1$ continues to be satisfied. This can be shown by substituting
$\alpha =|\alpha |e^{-i\psi }$ into Eq.~(6), multiplying by
$e^{i\psi }$, and taking the real part of the result, thus
obtaining
\begin{equation}
\partial |\alpha |/\partial t+(K/2)(|\alpha |^2-1)Re[re^{-i\psi
}]=0 \ .
\end{equation}
We see from Eq.~(9) that $\partial |\alpha |/\partial t=0$ at
$|\alpha |=1$.  Hence a trajectory of (6), starting with an
initial condition satisfying $|\alpha (\omega ,0)|<1$ cannot cross
the unit circle in the complex $\alpha $-plane, and we have
$|\alpha (\omega ,t)|<1$ for all finite time, $0\leq t < +\infty
$.

One way to motivate our ansatz, $f_n=\alpha ^n$, is to note that
the well-known stationary states of the Kuramoto
model\cite{Kuramoto,model}, both the incoherent state ($f=g/2\pi $
corresponding to $\alpha =0$) and the partially synchronized state
with $|r|=const.>0$, both conform to $f_n=\alpha ^n$. Thus one
view of the ansatz is that it specifies a family of distribution
functions that connect these two states in a natural way.

To proceed further, we now introduce another restriction on our
assumed form of $f$: we require that $\alpha (\omega ,t)$ be
analytically continuable from real $\omega $ into the complex
$\omega $-plane, that this continuation has no singularities in
the lower half $\omega$-plane, and that $|\alpha (\omega
,t)|\rightarrow 0$ as $Im(\omega )\rightarrow -\infty$. If these
conditions are satisfied for the initial condition, $\alpha
(\omega ,0)$, then they are also satisfied for $\alpha (\omega
,t)$ for $\infty
>t>0$. To see that this is so, we first note that for large
negative $\omega _i=Im(\omega )$, Eq.~(6) is approximately
$\partial \alpha /\partial t=-|\omega _i|\alpha $, and thus
$\alpha (\omega ,t)\rightarrow 0$ as $\omega _i\rightarrow
-\infty$ will continue to be satisfied if $\alpha (\omega
,0)\rightarrow 0$ as $\omega _i\rightarrow -\infty $. Next we note
from\cite{Coddington} that $\alpha (\omega ,t)$ is analytic in any
region of the complex $\omega $-plane for which $\alpha (\omega
,0)$ is analytic provided that the solution $\alpha (\omega ,t)$
to Eq.~(6) exists. To establish existence for $0\leq t < +\infty $
it suffices to show that the solution to Eq.~(6) cannot become
infinite at a finite value of $t$. This can be ruled out by noting
that our derivation of (9) with $\omega $ now complex carries
through except that there is now an addition term $-\omega
_i|\alpha |$ on the left hand side of the equation. Thus at
$|\alpha |=1$ we have $\partial |\alpha |/\partial t=\omega
_i|\alpha |<0$, and we conclude that, if $|\alpha (\omega ,0)|<1$
everywhere in the lower half complex $\omega $-plane, then
$|\alpha (\omega ,t)|<1$ for all finite time $0\leq t<+\infty $
everywhere in the lower half complex $\omega $-plane.

Regarding the initial condition $\alpha (\omega ,0)$, we note
that, if $|\alpha (\omega ,0)|\leq 1$ for $\omega $ real, if the
continuation $\alpha (\omega ,0)$ is analytic everywhere in the
lower half $\omega $-plane, and if the continuation satisfies
$|\alpha (\omega ,0)|\rightarrow 0$ as $\omega _i\rightarrow
-\infty$, then the continuation satisfies $|\alpha (\omega ,0)|<1$
everywhere in the lower half complex $\omega $-plane\cite{Cauchy}.
Examples of possible initial conditions are $k\exp (-i\omega c)$
with $Re(c)>0$ and $|k|\leq 1$, $k/(\omega -d)$ with $|k|\leq
Im(d)$, and $\int ^\infty _0 k(c)\exp (-i\omega c)dc$ with $\int
^\infty _0|k(c)|dc\leq 1$.

We can now specify the invariant manifold $M$ on which our
dynamics takes place. It is the space of functions of the real
variables $(\omega ,\theta )$ of the form given by Eq.~(8) where
$|\alpha (\omega ,t)|\leq 1$ for real $\omega $; $\alpha (\omega
,t)$ can be analytically continued from the real $\omega $-axis
into the lower half $\omega $-plane; and, when continued into the
lower half $\omega $-plane, $\alpha (\omega ,t)$ has no
singularities there and approaches zero as $\omega _i\rightarrow
-\infty $.

Now taking $g(\omega )$ to be Lorentzian
\[
g(\omega )=g_L(\omega )\equiv (\Delta /\pi )[(\omega - \omega
_0)^2+\Delta ^2]^{-1} \ ,
\]
we can do the $\omega $ integral in Eq.~(7) by closing the contour
by a large semicircle in the lower half $\omega$-plane. Writing
$g_L(\omega ) =(2\pi i)^{-1}[(\omega -\omega _0-i\Delta
)^{-1}-(\omega -\omega _0 +i\Delta )^{-1}]$, we see that the
integral is given by the residue of the pole at $\omega =\omega
_0-i\Delta $. By a change of variables $(\theta ,\omega
)\rightarrow (\theta -\omega _0t, (\omega -\omega _0)/\Delta )$,
we can, without loss of generality set $\omega _0=0$, $\Delta =1$.
Thus we obtain $r(t)=\alpha ^*(-i,t)$. Putting this result into
Eq.~(6) and setting $\omega =-i$, we obtain the nonlinear
evolution of the order parameter $r=\rho e^{-i\phi }$ $(\rho \geq
0$ and $\phi $ real):
\begin{equation}
d\rho /dt+\left( 1-\frac{1}{2}K\right) \rho +\frac{1}{2}K\rho ^3=0
\ ,
\end{equation}
and $d\phi /dt=0$.  Thus the dynamics is described by the single
real nonlinear, first order, ordinary differential equation,
Eq.~(10).  The solution of Eq.~(10) is,
\begin{equation}
\frac{\rho (t)}{R}=\left| 1+ \left[ \left( \frac{R}{\rho
(0)}\right) ^2-1\right] e^{(1-\frac{1}{2}K)t}\right|^{-1/2} \ ,
\end{equation}
where $R=|1-(2/K)|^{1/2}$.  We see that for $K<K_c=2$, the order
parameter goes to zero exponentially with increasing time, while
for $K>2$ it exponentially asymptotes to the finite value
$[1-(2/K)]^{1/2}$, in agreement with the known time-asymptotic
results for the case $g=g_L$ (e.g., see Ref.\cite{model}). Plots
of the nonlinear evolution of $\rho (t)$ are shown in Fig.~1.
\begin{figure}[h]
\epsfig{file=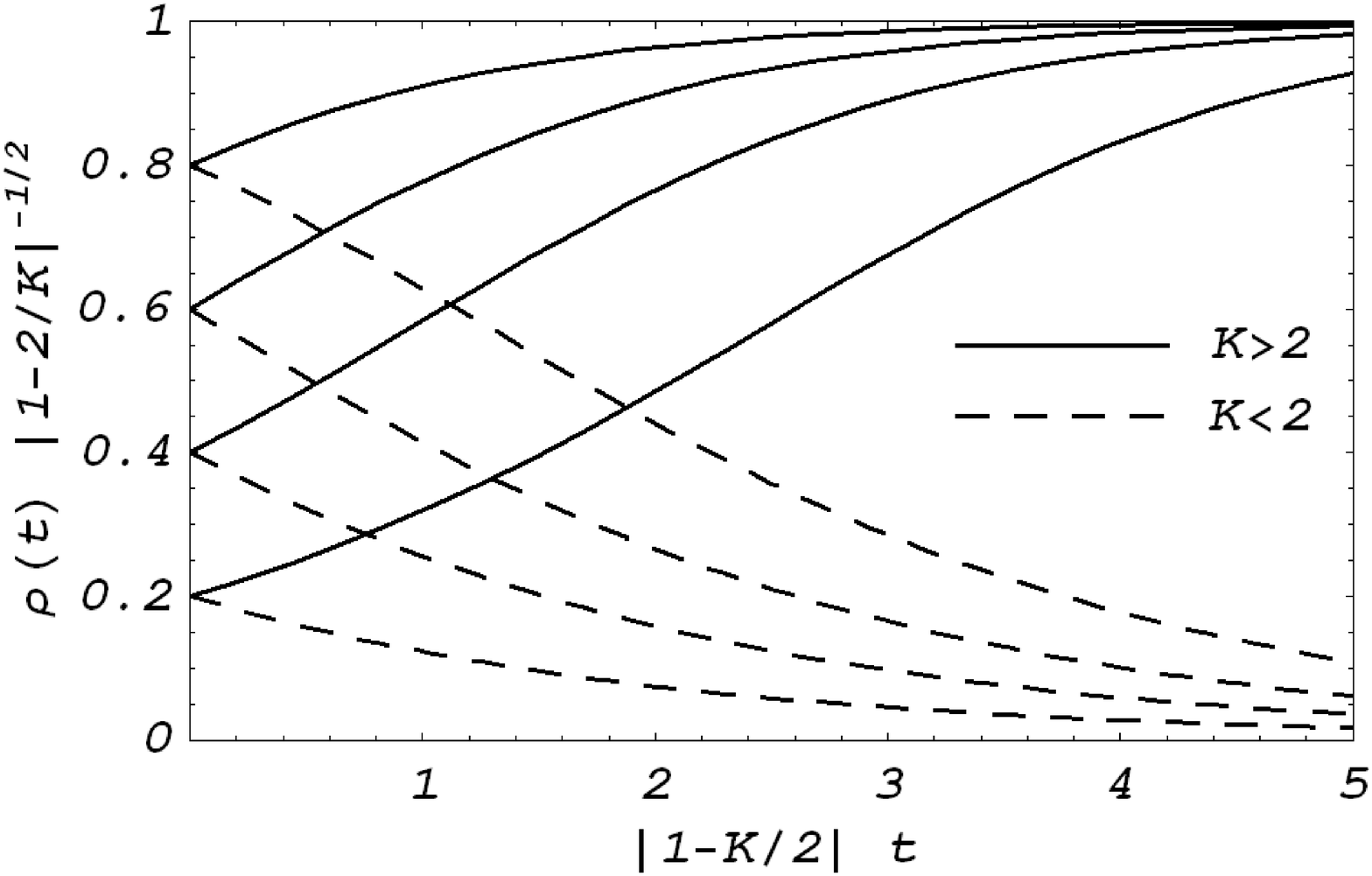,width=240pt} \caption{The order
parameter $\rho =|r|$ versus time. } \label{fig:Figure 1}
\end{figure}
 Linearization of Eq.~(10) yields an exponential damping
rate of $[1-(K/2)]$ for perturbations around $\rho =0$ for $K<2$,
which becomes unstable for $K>K_c=2$, at which point the stable
nonlinear equilibrium at $\rho =\sqrt{1-(2/K)}$ comes into
existence. For $K>K_c$ linearization of (10) around the
equilibrium $\rho =\sqrt{1-(2/K)}$ yields a corresponding
perturbation damping rate $[(K/2)-1]$. For $g=g_L$ the latter
damping rate can also be obtained from the recent stability
analyses of solutions of Eqs.~(4) and (5) \cite{Antonsen,Mirollo}.
We emphasize that our solution for $r(t)$ obeys two uncoupled
first order real ordinary differential equations (Eq.~(10) and
$d\phi /dt=0$), while the problem for $\alpha (\omega ,t)$
(Eqs.~(6) and (7)) is an infinite dimensional dynamical system
(i.e., to obtain $\alpha (\omega ,t)$ we need to specify an
initial {\it function} of $\omega $, $\alpha (\omega ,0)$). This
is further reflected by the fact that linearization of Eqs.~(6)
and (7) about their equilibria yields a problem with a continuous
spectrum of neutral modes\cite{Mirollo,Strogatz}. Thus the
microscopic dynamics in $M$ of the distribution function is
infinite dimensional, while the macroscopic dynamics of the order
parameter is low dimensional.

\section{IV. Generalizations}

{\it a. Other distributions $g(\omega )$}

So far we have restricted our discussion to the case of the
Lorentzian $g_L(\omega )$. We now consider
\[
g(\omega )=g_4(\omega )\equiv (\sqrt {2}/\pi )(\omega ^4+1)^{-1}\
, \]
 which decreases with increasing $\omega $ as $\omega ^{-4}$,
in contrast to $g_L(\omega )$ which decreases as $\omega ^{-2}$.
The distribution $g_4(\omega )$ has four poles at $\omega =(\pm
1\pm i)/\sqrt 2$. Proceeding as before, we apply the residue
method to the integral (7) to obtain
\[
r(t)=\frac{1}{2}[(1+i)r_1(t)+(1-i)r_2(t)] \ ,
\]
where
\[
r_{1,2}=\alpha ^*((\mp 1-i)/\sqrt{2},t)
\]
and $r_{1,2}(t)$ obey the two coupled nonlinear ordinary
differential equations,
\begin{equation}
dr_{1,2}/dt+(K/2)[r^*r^2_{1,2}-r]+[(1\mp i)/\sqrt{2}]r_{1,2}=0 \ .
\end{equation}
 Thus we obtain a system of two first order complex nonlinear
 differential
equations. Indeed, the above considerations can be applied to any
$g(\omega )$ that is a rational function of $\omega $ (i.e.,
$g(\omega )=P_1(\omega )/P_2(\omega )$ where $P_1(\omega )$ and
$P_2(\omega )$ are polynomials in $\omega $). The requirement that
$g(\omega )$ be normalizable $(\int g(\omega )d\omega =1)$ and
real puts restrictions on the possible $P_{1,2}(\omega )$; e.g.,
$P_2(\omega )$ must have even degree, $2m$, and all its roots must
come in complex conjugate pairs (it cannot have a root on the real
$\omega $ axis). Such a $g(\omega )$ has $m$ poles in the lower
half $\omega $-plane, and application of our method yields $m$
complex, first order ordinary differential equations for $m$
complex order parameters. For instance, for the example, $g(\omega
)=g_4(\omega )$, above, there are two poles in $Im(\omega )<0$,
namely, $\omega =(\pm 1-i)/\sqrt{2}$, and these two poles result
in the two order parameters $r_1$ and $r_2$.

{\it b. External driving}

We now consider the Kuramoto problem with an external drive,
Eq.~(2). Again taking the $N\rightarrow \infty$ limit for the
number of oscillators, we obtain
\begin{equation}
\frac{\partial f}{\partial t}+\frac{\partial}{\partial \theta
}\left\{ f\left[ (\omega -\Omega ) +\frac{1}{2i}(Kr+\Lambda
)e^{-i\theta }-\frac{1}{2i}(Kr+\Lambda )^*e^{i\theta
}\right]\right\}=0 \ ,
\end{equation}
with $r(t)$ given by Eq.~(5). In writing Eq.~(13), we have
utilized a change of variables $\theta \rightarrow \theta +\Omega
t$ to remove the $e^{i\Omega t}$ time dependence that would
otherwise appear multiplying the $\Lambda $ terms. Again assuming
that $g(\omega )$ is a Lorentzian with unit width $\Delta =1$
peaked at $\omega =\omega _0$, and proceeding as before, we obtain
the following equation for $r(t)$,
\begin{equation}
dr/dt+\frac{1}{2}\{(Kr+\Lambda )^*r^2-(Kr+\Lambda )\}+[1+i(\Omega
- \omega _0)]r=0 \ .
\end{equation}
Equilibria are obtained by setting $dr/dt=0$ in Eq.~(14).
Depending on parameters $(K,\Omega ,\Lambda )$, there are either
one or three such equilibria\cite{Antonsen}. Also, depending on
parameters, there may be an attracting limit cycle.  Whether the
equilibria are attractors for Eq.~(14) depends on their stability
which can be assessed by linearization around the equilibria. The
equilibria obtained for Eq.~(14) and their stability are the same
as obtained by the analysis of the full system (13) as performed
in Ref.\cite{Antonsen}. Furthermore, the bifurcations and
stability of the limit cycle are the same as numerically found in
Ref.\cite{Antonsen}. Thus, for this problem, it appears that the
important observable macroscopic dynamics is contained entirely
within the invariant manifold $M$.

{\it c. Communities of oscillators}

Turning now to the problem of coupled communities of Kuramoto
systems given by Eq.~(3), we introduce different Lorentzians for
each community,
\[
g^\sigma (\omega )=\pi ^{-1}[(\omega -\omega _\sigma )^2+\Delta
^2_\sigma ]^{-1}\ ,
\]
and proceed as before. We obtain a coupled
system of equations for the order parameter associated with each
community $\sigma $,
\begin{equation}
dr_\sigma /dt+(-i\omega _\sigma +\Delta _\sigma )r_\sigma
+\frac{1}{2}\sum ^s_{\sigma '=1}K_{\sigma \sigma '}[r^*_{\sigma
'}r^2_\sigma -r_{\sigma '}]=0 \ ,
\end{equation}
where $\sigma =1,2,\ldots ,s$. Thus we obtain $s$ complex coupled
differential equations where $s$ is the number of communities. We
conjecture that, for $s$ large enough [e.g., $s\geq 2$ or $3$] and
appropriate parameter values, there may be chaotic attracting
solutions for Eq.~(15). It would be particularly interesting to
see whether such solutions in $M$ are also attractors for the
macroscopic order-parameter behavior of the full system (3), e.g.,
by comparing numerical solutions of Eqs.~(3) and (15).

{\it d. Time-delayed coupling}

In applications time delay in the coupling between dynamical units
in a network is often present. For example, the propagation speed
of signals between units is finite (e.g., along axons in a neural
network), and there may also be an inherent response time of a
unit to information that it receives. Thus time delay has been
extensively studied in the context of networks of coupled systems,
and in particular for the case of coupled phase
oscillators\cite{Niebur,Yeung,Choi}. It has been found for such
systems that time delay can substantially modify the dynamics,
leading to a much richer variety of behaviors. In the context of
Eqs.~(1)--(3), for example, the response of oscillator $i$ at time
$t$ to input from oscillator $j$ is now related to the state
$\theta _j$ of oscillator $j$ at time $(t-\tau _{ji})$ where $\tau
_{ji}$ is the time delay for this interaction. Assuming that all
the delay times are the same, $\tau _{ji}=\tau $, independent of
$i$ and $j$, the quantities $\theta _j(t)$ appearing in the
summations in Eqs.~(1)--(3) must now be replaced by
$\theta_j(t-\tau )$. Again such a generalization can be
straightforwardly incorporated into our method. For example, for
the external drive problem (Eq.~(2) and Sec.~IVb) we have in place
of Eq.~(2),
\begin{equation}
\frac{d\theta _i(t)}{dt}=\omega _i+\frac{K}{N}\sum ^N_{i=1}\sin
[\theta _j(t-\tau )-\theta _i(t)]+\Lambda \sin [\Omega t-\theta
_i(t)] \ .
\end{equation}
Going to a rotating frame, $\theta '_i(t)=\theta _i(t)-\Omega t$,
$\omega '=\omega -\Omega $, Eq.~(16) becomes
\begin{equation}
\frac{d\theta ' _i(t)}{dt}=\omega ' _i+\frac{K}{N}\sum
^N_{i=1}\sin [\theta ' _j(t-\tau )-\theta ' _i(t)-\Omega \tau
]-\Lambda \sin \theta ' _i(t) \ .
\end{equation}
The summation in Eq.~(17) is
\begin{equation}
\frac{K}{N}Im\left\{ e^{-i[\theta '_i(t)+\Omega \tau ]}\sum
^N_{j=1}e^{i \theta '_j(t-\tau )}\right\} =K Im\left\{
e^{-i[\theta '_i(t)+\Omega \tau ]}r(t-\tau )\right\} \ .
\end{equation}
Thus, to include delay, it suffices to replace
 the term $[Kr(t)+\Lambda ]$ in Eqs.~(13) and (14) by $[Ke^{-i\Omega \tau }r(t-\tau )+\Lambda ]$. E.g.,
making this substitution in Eq.~(14) and setting $\Lambda =0$,
$\Omega =\omega _0$ yields the following first order
delay-differential equation for the order-parameter of the
standard Kuramoto model with coupling delay,
\begin{equation}
\frac{dr(t)}{dt}-\frac{K}{2}\left[ e^{-i\omega _{0} \tau }r(t-\tau
)-e^{i\omega _0\tau }r^*(t-\tau )(r(t))^2\right] +r(t)=0\ ,
\end{equation}
which returns Eq.~(10) for $\tau \rightarrow 0$. We note that our
reduced descriptions with delay (e.g., Eq.~(19)) are (in contrast
to Eqs.~(10), (14) and (15)) now infinite dimensional dynamical
systems. For small $|r|$, linearizing Eq.~(19) about the
incoherent state $(r=0)$, and setting $r\sim e^{st}$ yields a
dispersion relation for $s$,
\begin{equation}
s+1=(K/2)\exp [-(s+i\omega _0)\tau ] \ ,
\end{equation}
in agreement with Ref.~\cite{Yeung}. In addition, steady
synchronized states can be found (as in Ref.~\cite{Choi}) by
setting $r(t)=r_0e^{i\eta t}$ in Eq.~(19) and solving the result,
\begin{equation}
i\eta -\frac{K}{2}\left[ e^{-i(\omega _0+\eta )\tau }
-r_0^2e^{i(\omega _0+\eta )\tau }\right]+1=0 \ ,
\end{equation}
for the real constants $\eta $ and $r_0$. Furthermore, through
linearization of Eq.~(19) about $r=r_0e^{i\eta t}$, our
formulation can be used to study the previously unaddressed
problem of assessing the stability of the steady synchronized
states, Eq.~(21).

 {\it e. The Millennium bridge problem, Ref.\cite{Eckhardt}}

Another example is that of the observed oscillation of London's
Millennium Bridge induced by the pacing phase entrainment of
pedestrians walking across the bridge as modeled by Eqs.~(52) and
(53) of the paper by B. Eckhardt et al.\cite{Eckhardt}. In that
case, assuming a Lorentzian distribution of natural pacing
frequencies for the pedestrians, one can use the method given in
our paper to obtain an ordinary differential equation for the
mechanical response of the bridge coupled to another ordinary
differential equation for the order parameter describing the
collective state of the pedestrians.

\section{V. Discussion and Conclusion}

Low dimensional descriptions of the classical Kuramoto problem
(Eq.~(1)) have been previously attempted.  An early such attempt
was made by Kuramoto and Nishikawa\cite{Kuramoto86} who used a
heuristic approach resulting in an integral equation for $r(t)$.
On the basis of their work they predict that for small $|r(0)|$
the order-parameter $r(t)$ initially grows (decays) exponentially
in time for $K>K_c$ $(K<K_c)$ (later shown rigorously and
quantitatively in Ref.~\cite{Strogatz}). Crawford\cite{Crawford},
using center manifold theory, obtains (Eq.~(108) of
Ref.~\cite{Crawford}) an equation of the form $d\rho /dt
=a(K-K_c)\rho +b\rho ^3+O(\rho ^5)$ for $K$ near $K_c$.  Another
work of interest is that of Watanabe and Strogatz\cite{Watanabe}
who consider the case where all oscillators have the same
frequency for both finite and infinite $N$. By use of a nonlinear
transformation of the phase variables $\theta _i(t)$, these
authors show that the dynamics reduces to a solution of three
coupled first order ordinary differential equations. Thus, while
macroscopic behavior of order-parameter dynamics has been
previously addressed for the standard Kuramoto problem, it has,
until now, never been demonstrated fully (e.g., without the
restriction of \cite{Crawford} to small amplitude, or the
restriction of \cite{Watanabe} to identical frequencies). Our
paper does this and also demonstrates that our technique can be
usefully applied to a host of other important related problems.

 Our work also suggests other
future lines of study. For example, can any rigorous results be
obtained relevant to whether our macroscopic order-parameter
attractors obtained by considering $f$ in the manifold $M$ have
general validity\cite{24}? Are there interesting qualitative
differences between the behavior for Lorentzian $g(\omega )$ as
compared to other monotonic symmetric oscillator distribution
functions $g(\omega )$? What other systems, in addition to those
discussed in Sec.~IV, can our method be applied to?

We thank B. R. Hunt, M. Girvan, R. Faghih and J. Platig for very
useful interactions. This work was supported by the ONR
(N00014-07-1-0734) and by the NSF (PHY0456249).

\newpage

\section*{Figure Captions}
\newcounter{fig}
\begin{list}{\bfseries\upshape Figure \arabic{fig}:}
{\usecounter{fig}
\setlength{\labelwidth}{2cm}\setlength{\leftmargin}{2.6cm}
\setlength{\labelsep}{0.5cm}\setlength{\rightmargin}{1cm}
\setlength{\parsep}{0.5ex plus0.2ex minus 0.1ex}
\setlength{\itemsep}{0ex plus0.2ex} \slshape}

\item The order parameter $\rho =|r|$ versus time.
\end{list}

\begin{references}
\bibitem{Cauchy} This follows by noting that the Cauchy-Riemann
conditions imply that the complex function $\alpha (\omega
_r+i\omega _i,0)$ satisfies Laplace's equation in $\omega $,
$\nabla ^2_\omega \alpha (\omega _r+i\omega _i,0)=0$, where
$\nabla ^2_\omega =\partial ^2/\partial \omega ^2_r+\partial
^2/\partial \omega _i^2$. Setting $\alpha =|\alpha |e^{-i\psi }$,
this gives $\nabla ^2_\omega |\alpha |=|\alpha |[(\partial \psi
/\partial \omega _r)^2+(\partial \psi /\partial \omega _i)^2]$.
Since $\nabla ^2_\omega |\alpha |\geq 0$ in $\omega _i<0$,
$|\alpha |$ cannot assume a maximum anywhere interior to the
lower-half complex $\omega $-plane. Hence the maximum value of
$|\alpha |$ in $\omega _i\leq 0$ must occur on $\omega _i=0$,
i.e., on the real $\omega $-axis.
\bibitem{24} While the time-asymptotic behavior of the
order-parameter obtained from the dynamics of the full system is
seen to correspond to the attractors of the reduced systems (10)
and (13)  (where (13) is treated in [10]), we emphasize that $M$
need not be attracting for the {\it microscopic} state $f(\omega
,\theta ,t)$. This can be simply seen in the case of the Kuramoto
problem (4), for which $f(\omega ,\theta ,t)=[g(\omega )/2\pi ] \{
1+\sum _m\beta _m(\omega )\exp [im(\theta -\omega t)]\}$ and
$r(t)=0$ for all $t$ is a solution for $\beta _{\pm 1}(\omega )=0$
and {\em any} choice of perturbations $\beta _m(\omega )$,
$|m|\geq 2$.
\end{references}
\end{document}